\documentclass[conference]{IEEEtran}
\usepackage{latexsym}
\usepackage{graphicx}
\usepackage{amsfonts,amssymb,amsmath}
\usepackage{ctable} 
\usepackage{mathtools, cuted}
\usepackage{hyperref}
\usepackage[ruled,linesnumbered,noend]{algorithm2e}
\usepackage[T1]{fontenc}
\usepackage{cite}
\usepackage{subcaption}
\usepackage{comment}
\usepackage{diagbox}

\usepackage[a4paper, left=0.662in, right=0.662in, top=0.7in]{geometry}
\usepackage{amsthm}
\usepackage{overpic}
\usepackage{steinmetz}
\usepackage{array}
\usepackage{url}
\usepackage{color}
\usepackage{algorithmicx}
\usepackage{algcompatible}

\usepackage{xcolor}
\usepackage{soul}

\theoremstyle{plain}

\def\Htran{\mbox{\tiny $\mathrm{H}$}}
\def\Ttran{\mbox{\tiny $\mathrm{T}$}}

\IEEEoverridecommandlockouts

\title{Machine Learning-Based Near-Field Localization in Mixed LoS/NLoS Scenarios}
\author{Parisa Ramezani\textsuperscript{1}, Seyed Jalaleddin Mousavirad\textsuperscript{2}, Mattias O'Nils\textsuperscript{2}, and Emil Bj\"{o}rnson\textsuperscript{1}\\
\IEEEauthorblockA{\textit{\textsuperscript{1}Department of Computer Science, KTH Royal Institute of Technology, Stockholm, Sweden} \\
\textit{\textsuperscript{2}Department of Computer and Electrical Engineering, Mid Sweden University, Sundsvall, Sweden
} \\  Email: \{parram,\,emilbjo\}@kth.se\,,\{seyedjalaleddin.mousavirad,\,mattias.onils\}@miun.se}%
\thanks{This work was supported by the FFL18-0277 grant and SUCCESS project (FUS21-0026), funded by the Swedish Foundation for Strategic Research. This research also was funded by the Swedish Knowledge Foundation through the Research Profile NIIT.}
}

\begin{document}
\maketitle

\begin{abstract}
The conventional MUltiple SIgnal Classification (MUSIC) algorithm is effective for angle-of-arrival estimation in the far-field and can be extended for full source localization in the near-field. However, it suffers from high computational complexity, which becomes especially prohibitive in near-field scenarios due to the need for exhaustive 3D grid searches. This paper presents a machine learning-based approach for 3D localization of near-field sources in mixed line-of-sight (LoS)/non-LoS scenarios. A convolutional neural network (CNN) learns the mapping between the eigenvectors of the received signal's covariance matrix at the anchor node and the sources' 3D locations. The detailed description of the proposed CNN model is provided. The effectiveness and time efficiency of the proposed CNN-based localization approach is corroborated via numerical simulations.
\end{abstract}

\section{Introduction}
Wireless localization has long been a fundamental problem in array signal processing with applications in various fields such as sonar, radar, indoor positioning systems, cellular communications, search and rescue operations, and wireless sensor networks \cite{Wax1983,Malioutov,Patwari-locating}. 
With the advent of new technologies such as autonomous systems and smart cities, and the need for accurate location information for seamless connectivity in wireless communications, localization has become a cornerstone for next-generation wireless networks. 

Among numerous localization techniques, MUltiple SIgnal Classification (MUSIC) is one of the most prominent ones due to its high resolution and accuracy, ability to locate multiple sources simultaneously, and robustness against noise. This technique was initially developed for angle-of-arrival (AoA) estimation \cite{Schmidt-MUSIC} and later extended to near-field localization where both the azimuth AoA and range of the near-field sources were estimated to locate the sources \cite{Huang-NearField}. However, MUSIC requires a high-resolution grid to accurately localize near-field sources, which leads to a high computational complexity. 
This computational burden becomes prohibitive in 3D localization scenarios using a uniform planar array (UPA) as the anchor node, where azimuth AoA, elevation AoA, and range must be estimated. In such a case, the
complexity grows cubically with the grid resolution per dimension. Some techniques have been developed lately for reducing the complexity of 3D MUSIC by decoupling the AoA and range estimation problems \cite{Alva-parametric,Parisa-modified}. These techniques relax the 3D spectral search by transforming it into a 2D search and multiple 1D searches; however, the 2D search still possesses a relatively high computational overhead, especially for a fine grid resolution. 

Recently, machine learning, and specifically its specialized branch of deep learning (DL), has gained interest as a promising approach to address the challenges faced by the MUSIC algorithm. Reference \cite{fuchs2019single} proposes a multi-layer perceptron neural network architecture to estimate the azimuth AoAs of two sources using one snapshot of received signals. In this work, an artificial neural network (ANN) is trained using simulated and real-world data and the spatial covariance matrix is set as the ANN input. 
One of the most popular DL models is the convolutional neural network (CNN), which is designed to process structured data. CNNs use layers of filters to extract important features automatically and reduce the need for manual feature selection \cite{Deep_Image01,CNN_Jalal}. A DL framework is introduced in \cite{deepMUSIC} with multiple CNNs, each dedicated to a sub-region of the angular spectrum. The covariance matrix is taken as the input and the CNNs are trained to predict the spectrum for their assigned sub-region. 
Another CNN-based approach is developed in \cite{alteneiji2021angle} in which a multi-path propagation environment is considered, the eigenvectors of the covariance matrix are used as the input, and the CNN outputs the azimuth AoA of two sources. Reference \cite{mylonakis3d} considers a $4\times 4$ UPA as the anchor node and develops a CNN model for estimating both azimuth and elevation AoAs of the sources. 

This paper introduces a machine learning-based approach for near-field localization in mixed line-of-sight (LoS)/non-LoS (NLoS) environments. 
The models proposed in \cite{fuchs2019single,deepMUSIC,alteneiji2021angle,mylonakis3d} are not applicable to the 3D localization framework considered in this paper, as they are limited to estimating the sources' AoAs without providing range information.
We implement a CNN architecture that processes the eigenvectors of the received signal's covariance matrix to directly estimate the 3D coordinates of near-field sources in the $x$-$y$-$z$ plane. We validate the effectiveness of the proposed approach by comparing it with the MUSIC algorithm, demonstrating that our method achieves comparable or better performance while significantly reducing computational overhead and runtime. 

\section{System Model}
Assume that an anchor node in the form of a UPA with $N = N_y \times N_z$ antennas, located in the $y$-$z$ plane, aims to estimate the location of $K$ near-field signal sources. At discrete time slot $t$, the received signal at the anchor node is given by 
\begin{equation}
\mathbf{x}(t) = \sum_{k=1}^K \mathbf{h}_k s_k(t) + \mathbf{n}(t),    
\end{equation}
where $\mathbf{x}(t) = [x_1(t),\ldots,x_N(t)]^{\Ttran}$ is the signal received by the $N$ antennas of the anchor node, $s_k(t)$ is the random signal from the $k$th source, which is unknown to the receiver, and $\mathbf{n}(t) = [n_1(t),\ldots,n_N(t)]^{\Ttran}$ is the additive independent complex Gaussian noise. The signals are assumed to be uncorrelated across different sources and different time slots.
Furthermore, $\mathbf{h}_k \in \mathbb{C}^{N}$ represents the channel between the $k$th source and the anchor node which consists of both LoS and NLoS components. Specifically, it is modeled by Rician fading as $\mathbf{h}_k = \sqrt{\frac{\kappa}{\kappa+1}} \mathbf{h}_{k,\mathrm{LoS}} + \sqrt{\frac{1}{\kappa + 1}} \mathbf{h}_{k,\mathrm{NLoS}}$,
where $\kappa$ denotes the Rician factor,  $\mathbf{h}_{k,\mathrm{LoS}} = \mathbf{a}(\varphi_k,\theta_k,\bar{r}_k)$ is the LoS channel with $\mathbf{a}(\varphi_k,\theta_k,\bar{r}_k)$ being the array response vector for the signal received from the $k$th source at the anchor node. $\varphi_k$, $\theta_k$, and $\bar{r}_k$ denote the azimuth AoA, elevation AoA, and range of the $k$th source with respect to the reference antenna at the anchor node. 
Assuming that the reference antenna is located at $(0,0,0)$, the array response vector for the $k$th source is given by 
\begin{equation}
\mathbf{a}(x_k,y_k,z_k) = \left[1,\ldots,e^{j\frac{2\pi}{\lambda}(\bar{r}_k - r_k^n )},\ldots, e^{j\frac{2\pi}{\lambda}(\bar{r}_k - r_k^N )} \right]^{\Ttran},    
\end{equation}
where $\lambda$ is the wavelength and
\begin{equation}
\label{eq:distance}
  r_k^n = \sqrt{x_k^2 + \left(y_k - n_y d_y\right)^2 + \left(z_k - n_z d_z\right)^2}  
\end{equation} is the distance between the $k$th source and the $n$th antenna. $(x_k,y_k,z_k)$ represents the coordinate of the $k$th source with $x_k = \bar{r}_k \cos(\varphi_k)\cos(\theta_k)$, $y_k = \bar{r}_k \sin(\varphi_k)\cos(\theta_k)$, $z_k = \bar{r}_k \sin(\theta_k)$ .
Furthermore, $n_y \in \{1,\ldots,N_y\}$ and $n_z \in \{1,\ldots,N_z\}$ indicate the horizontal and vertical indices of the $n$th antenna, and $d_y$ and $d_z$ are the inter-antenna spacing in the respective dimensions. 
$\mathbf{h}_{k,\mathrm{NLoS}} \in \mathbb{C}^N$ represents the NLoS component consisting of scattered multi-path channels based on correlated Rayleigh fading, as described in \cite{demir2024spatial}. 

\section{Multiple Signal Classification}
The covariance matrix of the received signal at the anchor node is expressed as 
\begin{equation}
\label{eq:eigendecomposition}
  \mathbf{R} = \mathbb{E}\{\mathbf{x}(t) \mathbf{x}^{\Htran}(t)\} = \mathbf{U}\mathbf{D}\mathbf{U}^{\Htran},  
\end{equation}
where the term in the right-hand-side represents the eigendecomposition of $\mathbf{R}$. The diagonal entries of $\mathbf{D}$ contain the eigenvalues of the covariance matrix and the columns of $\mathbf{U}$ correspond to unit-length eigenvectors.
The MUSIC algorithm utilizes the orthogonality between the signal and noise subspaces of the received signal to estimate the location of the sources. Specifically, we further decompose \eqref{eq:eigendecomposition} into its signal and noise subspaces as $\mathbf{U}\mathbf{D}\mathbf{U}^{\Htran} = \mathbf{U}_s\mathbf{D}_s\mathbf{U}_s^{\Htran} + \mathbf{U}_n\mathbf{D}_n\mathbf{U}_n^{\Htran}$, where the diagonal matrices $\mathbf{D}_s$ and $\mathbf{D}_n$ has the $K$ highest and $N-K$ smallest eigenvalues of $\mathbf{R}$ on their diagonal, and $\mathbf{U}_s$ and $\mathbf{U}_n$ contain the corresponding eigenvectors. The MUSIC spectrum is defined as 
\begin{equation}
  S(\varphi,\theta,\bar{r}) = \frac{1}{\mathbf{a}^{\Htran}(\varphi,\theta,\bar{r})\mathbf{U}_n \mathbf{U}^{\Htran}_n \mathbf{a}(\varphi,\theta,\bar{r})},  
\end{equation}
and a 3D grid search is performed across the parameter space to identify spectral peaks that correspond to the estimated source locations in 3D space. This method would be asymptotically optimal if the grid were continuous. However, computing the MUSIC spectrum on a 3D grid presents a high computational burden in practice and makes it challenging to utilize the algorithm for real-time 3D localization applications. 

In practice, the covariance matrix is estimated by averaging over a finite number of snapshots as 
\begin{equation}
\label{eq:covariance_matrix_est}
  \hat{\mathbf{R}} =\frac{1}{T} \sum_{t=1}^{T}\mathbf{x}(t) \mathbf{x}^{\Htran}(t).  
\end{equation}

\section{Machine Learning-Based Near-Field Localization}
\label{sec:approach}
In this section, we present a convolutional neural network (CNN)-based method for localizing near-field sources. Specifically, we employ 2D-CNNs due to the two-dimensional spatial structure of the input eigenvector matrix. Unlike multilayer perceptrons or 1D-CNNs, 2D-CNNs are more effective at capturing spatial dependencies while requiring fewer parameters and offering greater accuracy.
The setup used for training and testing includes an anchor node equipped with $N = 128$ antennas with $N_y = 16$ and $N_z = 8$, and the inter-antenna spacing of $d_y = d_z = \lambda/2$, where $\lambda = 0.1$\,m. 
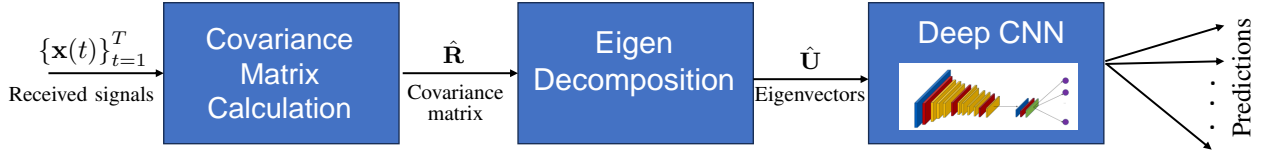
\begin{figure*}[t]
\centering
\begin{overpic}[width=0.95\textwidth]{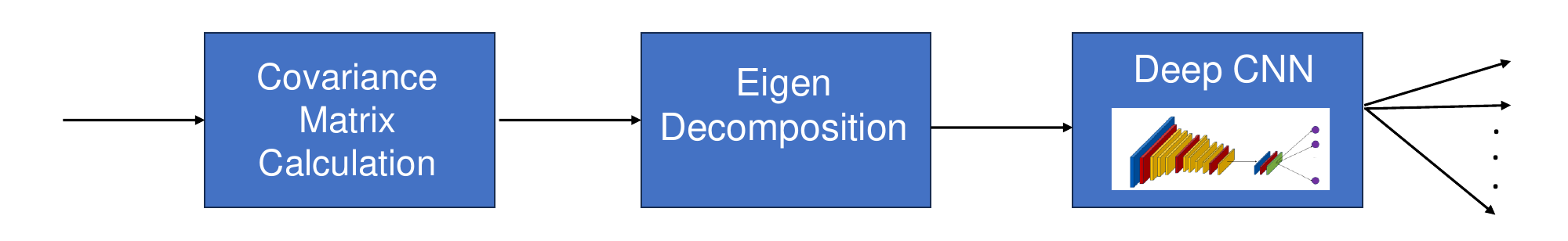}%
\put(0.9,5.5){\footnotesize Received signals}%
\put(3.2,8.9){$\left\{\mathbf{x}(t)\right\}_{t=1}^T$}%
\put(59.5,5.5){\footnotesize Eigenvectors}%
\put(63,8){\small $\hat{\mathbf{U}}$}%
\put(32.2,5.6){\footnotesize Covariance}%
\put(33.8,4){\footnotesize matrix}%
\put(35,8.5){\small $\hat{\mathbf{R}}$}%
\put(97,2.6){\rotatebox{90} {Predictions}}%
\end{overpic}
\caption{The general block diagram of the proposed approach. }
\label{fig:approach}
\vspace{-4mm}
\end{figure*}

Fig.\,\ref{fig:approach} shows the general block diagram of our proposed approach. First, the covariance matrix is calculated based on the received snapshots as in \eqref{eq:covariance_matrix_est}. The eigendecomposition is then performed on the covariance matrix 
as $\hat{\mathbf{R}} = \hat{\mathbf{U}} \hat{\mathbf{D}} \hat{\mathbf{U}}^{\Htran}$ and the matrix of eigenvectors $\hat{\mathbf{U}}$ is used as the input to the CNN as in \cite{alteneiji2021angle}. 
Each element of $\hat{\mathbf{U}}$ is a complex number.  
To enable processing within a real-valued DL framework, we decompose  $\hat{\mathbf{U}}$ into its real and imaginary components, forming a new representation: $\mathbf{X}_{\text{real}} = \Re(\hat{\mathbf{U}})$,  $\mathbf{X}_{\text{imag}} = \Im(\hat{\mathbf{U}})$.
The final input to the CNN is constructed as a multi-channel real-valued tensor as

\begin{equation}
    \mathbf{X}_{\text{input}} = \begin{bmatrix} \mathbf{X}_{\text{real}} \\ \mathbf{X}_{\text{imag}} \end{bmatrix} = \begin{bmatrix} \Re(\hat{\mathbf{U}}) \\ \Im(\hat{\mathbf{U}}) \end{bmatrix} \in \mathbb{R}^{2 \times N \times N}.
\end{equation}

In the following, we first provide a detailed explanation of the individual components of the model. Subsequently, we describe the overall architecture and its design rationale.
\subsection{Model Components} 

\subsubsection{Convolutional Blocks}
Convolutional blocks form the feature extraction backbone of the model. Each block consists of multiple convolutional layers with \( 3 \times 3 \) kernels, batch normalization, and ReLU activation functions. Let the input feature map at layer \( l \) be denoted as \( \mathbf{X}^{(l)} \), and the convolutional operation be represented as
\begin{equation}
    \mathbf{X}^{(l+1)} = \omega\left(\text{BN} \left( \mathbf{W}^{(l)} * \mathbf{X}^{(l)} + \mathbf{b}^{(l)} \right) \right),
\end{equation}
where \( \mathbf{W}^{(l)} \) and \( \mathbf{b}^{(l)} \) are the weight and bias parameters, \( * \) denotes the convolution operation, \( \text{BN}(\cdot) \) represents batch normalization, and \( \omega(\cdot) \) is the ReLU activation function. Max-pooling layers are used in the initial blocks to reduce spatial dimensions, while adaptive average pooling in the final block ensures a fixed output size. The feature depth increases progressively as
$32 \rightarrow 64 \rightarrow 128 \rightarrow 256$, capturing hierarchical patterns in the input data.

\subsubsection{Batch Normalization}
Batch normalization is applied after each convolutional and fully connected layer to standardize the inputs, stabilize training, and accelerate convergence. Given an input feature map \( \mathbf{X} \), batch normalization is defined as
\begin{equation}
    \text{BN}(\mathbf{X}) = \gamma \frac{\mathbf{X} - \mu}{\sqrt{\sigma^2 + \epsilon}} + \beta,
\end{equation}
where \( \mu \) and \( \sigma^2 \) are the batch mean and variance, \( \gamma \) and \( \beta \) are learnable parameters, and \( \epsilon \) is a small positive constant for numerical stability.

\subsubsection{Residual Connections}
Residual connections are incorporated by stacking multiple convolutional layers within each block. Given an input \( \mathbf{X}^{(l)} \), the residual mapping is formulated as:
\begin{equation}
    \mathbf{X}^{(l+1)} = \mathbf{X}^{(l)} + \mathcal{F}(\mathbf{X}^{(l)}, \mathbf{W}^{(l)}),
\end{equation}
where \( \mathcal{F}(\cdot) \) represents the residual function (typically a series of convolutions, batch normalization, and activation layers). 

\subsubsection{Fully Connected Layers}
The fully connected (FC) layers process the flattened feature maps from the convolutional blocks. Let the input to the first FC layer be \( \mathbf{z}_1 \). The FC layers progressively reduce dimensionality as $1024 \rightarrow 512 \rightarrow 256$.
Each layer transformation is defined as
\begin{equation}
    \mathbf{z}_{i+1} = \phi \left( \text{BN} (\mathbf{W}_{i} \mathbf{z}_i + \mathbf{b}_i) \right),
\end{equation}
where \( \mathbf{W}_i \) and \( \mathbf{b}_i \) are the weight and bias parameters, $\phi$ is the activation function, and dropout is applied to prevent overfitting.

\subsubsection{Output Layer}
The final FC layer produces a $3K$-dimensional output vector \( \mathbf{q} \in \mathbb{R}^{3K} \), where $3K$ is the number of target parameters - three parameters for the 3D location of each of the $K$ sources. It is given by 
\begin{equation}
    \mathbf{q} = \text{softmax} (\mathbf{W}_o \mathbf{z}_f + \mathbf{b}_o),\end{equation}
where \( \mathbf{W}_o \) and \( \mathbf{b}_o \) are the parameters of the output layer, and  $\mathbf{z}_f$ represents the output of the last FC layer before the final output layer.

\subsection{General Structure}

Based on the above-mentioned components, our proposed model is a CNN designed for regression tasks. It consists of a sequence of convolutional blocks for hierarchical feature extraction, followed by FC layers for prediction. Residual connections are incorporated into the convolutional blocks to improve gradient flow and enable deeper architectures.

\subsubsection{Convolutional Blocks}

The model consists of four convolutional blocks, each designed to progressively extract hierarchical features from the input data. The blocks share a similar structure but differ in the number of filters and the use of pooling operations. The general structure of each block is as follows:

\textbf{1. Convolutional Layers}:  
Each block contains multiple convolutional layers with $3 \times 3$ kernels, batch normalization, and ReLU activation. The number of filters increases progressively across the blocks:
\begin{itemize}
    \item Block 1: 2 input channels $\rightarrow$ 32 filters.
    \item Block 2: 32 filters $\rightarrow$ 64 filters.
    \item Block 3: 64 filters $\rightarrow$ 128 filters.
    \item Block 4: 128 filters $\rightarrow$ 256 filters.
\end{itemize}

\textbf{2. Pooling Operations}:  
\begin{itemize}
    \item Blocks 1 and 2: A max-pooling layer with a $2 \times 2$ kernel and stride 2 is applied to reduce spatial dimensions.
    \item Block 3: No pooling is applied; the block focuses on feature extraction through convolutional layers.
    \item Block 4: An adaptive average pooling layer is used to produce fixed-size feature maps.
\end{itemize}

\textbf{3. Residual Connections}:  
Residual connections are incorporated within the convolutional blocks by stacking multiple convolutional layers. This ensures efficient gradient propagation and enables the training of deeper architectures.

The operations in each block are summarized in Table~\ref{tab:block_ops}.

\begin{table*}[h]
\centering
\small
\caption{Block Operations. Abbreviations: $\text{Conv2D}$ (2D convolution), $\text{BN}$ (batch normalization), $\text{ReLU}$ (rectifier linear unit), $\text{MaxPool}$ (max-pooling), $\text{AAP}$ (adaptive average pooling).}
\label{tab:block_ops}
\begin{tabular}{cl}
\hline
\textbf{Block} & \textbf{Operations} \\ \hline
1, 2 & $\text{Conv2D} \rightarrow \text{BN} \rightarrow \text{ReLU} \rightarrow \text{Conv2D} \rightarrow \text{BN} \rightarrow \text{ReLU} \rightarrow \text{MaxPool}$ \\
3 & $\text{Conv2D} \rightarrow \text{BN} \rightarrow \text{ReLU} \rightarrow \text{Conv2D} \rightarrow \text{BN} \rightarrow \text{ReLU} \rightarrow \text{Conv2D} \rightarrow \text{BN} \rightarrow \text{ReLU}$ \\
4 & $\text{Conv2D} \rightarrow \text{BN} \rightarrow \text{ReLU} \rightarrow \text{Conv2D} \rightarrow \text{BN} \rightarrow \text{ReLU} \rightarrow \text{Conv2D} \rightarrow \text{BN} \rightarrow \text{ReLU} \rightarrow \text{AAP}$ \\ \hline
\end{tabular}
\end{table*}

\subsubsection{Fully Connected Layers}

The output of the final convolutional block is flattened into a 1D vector and processed through multiple FC layers. The FC layers are structured as follows:
\begin{itemize}
    \item \textbf{FC1} reduces the dimensionality from the flattened input size to 1024 neurons.
    \item \textbf{FC2} reduces the dimensionality to 512 neurons.
    \item \textbf{FC3} reduces the dimensionality to 256 neurons.
\end{itemize}

Each FC layer applies batch normalization, ReLU activation, and dropout for regularization. The final FC layer maps the 256-dimensional features to an output vector of size $3K$. This output vector represents the model's predictions, i.e., the 3D location of the $K$ sources. 

\section{Results}
In this section, we validate the effectiveness of the proposed CNN-based localization algorithm by comparing it with the 3D MUSIC algorithm. 
The goal is to localize $K = 3$ sources located in the radiative near-field region of the anchor node. 
The training and test datasets are generated through random realizations of the sources' locations. Specifically, for each sample in the datasets, the azimuth AoAs, elevation AoAs, and ranges of the sources are independently drawn from uniform distributions
 $\varphi_k \in \mathcal{U}[-\pi/3, \pi/3]$, $\theta_k \in \mathcal{U}[-\pi/3, \pi/3]$, and $\bar{r}_k \in \mathcal{U}[2D, d_{\mathrm{FA}}/4]$. 
 Here, $D$ is the array aperture length and $2D$ represents the distance beyond which amplitude variations over the array are negligible \cite{emil_primer}. Furthermore, $d_{\mathrm{FA}} = 2D^2/\lambda$ is the Fraunhofer array distance beyond which phase variations over the array are also negligible. We assume that the signal-to-noise ratio is $0$\;dB per antenna for all sources.
 The details of the deep model are provided in Section\,\ref{sec:approach} and the values of other hyperparameters are listed in Table\,\ref{tab:hyperparameters}. For the MUSIC algorithm, we used a grid size of 200 for all three parameters (azimuth angle, elevation angle, and range), which corresponds to a $0.6^\circ$ step size for angles and a $1$\,cm step size for range. The 3D coordinates of the sources are then found from the estimated AoAs and ranges. 
The experiments were conducted on a laptop running Windows 11, equipped with a 13th Gen Intel\textsuperscript{\textregistered} Core\textsuperscript{TM} i9-13900H processor (2.60 GHz), 64 GB of RAM, and an NVIDIA GeForce RTX 4090 Laptop GPU.
\vspace{-2mm}
\begin{table}[h]
 \centering
 \label{tab:hyperparameters}
     \caption{Hyperparameter values.}
    \centering
    \begin{tabular}{lc}
        \toprule
        \textbf{Hyperparameter} & \textbf{Value} \\
        \midrule
        Learning rate & $0.0001$ \\
        Weight decay & $0.01$ \\
        Optimizer & AdamW \\
        Training size & $20000$ \\
        Dropout rate & $0.3$ \\
         Number of epochs & $1200$ \\
          Batch size & 32 \\
          Loss function & MSE \\
        \bottomrule
    \end{tabular}
    \label{tab:hyperparameters}
    \vspace{-1mm}
\end{table}

We first evaluate the root mean square error (RMSE) of the location estimate for the proposed approach and the MUSIC algorithm, which is calculated as 
\begin{equation}
\begin{small}
\begin{aligned}
 &\mathrm{RMSE} =\\
 &\sqrt{\frac{1}{3KL} \sum_{k=1}^K\sum_{l = 1}^L(x_{k,t} - \hat{x}_{k,l})^2 + (y_{k,l} - \hat{y}_{k,l})^2 + (z_{k,l} - \hat{z}_{k,l})^2}, 
 \end{aligned}
 \end{small}
\end{equation}
where $x_{k,l}$, $y_{k,l}$, and $z_{k,l}$ represent the coordinates of the $k$th source for the $l$th test sample, $\hat{(\cdot)}$ indicates the estimated value, and the size of the test set is $L = 50$.
\begin{figure}
    \centering
    \includegraphics[width=\linewidth]{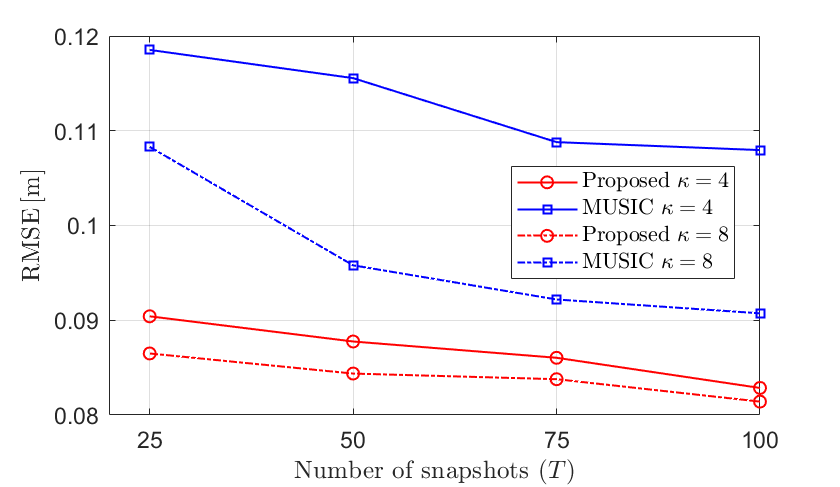}
    \caption{RMSE for the proposed approach and the MUSIC algorithm as a function of the number of snapshots ($T$).}
    \label{fig:rmse}
     \vspace{-4mm}
\end{figure}

Fig.\,\ref{fig:rmse} shows RMSE as a function of number of snapshots for Rician factors $\kappa = 4$ and $\kappa = 8$. The RMSE improves with the number of snapshots because with more snapshots, the covariance matrix can be better estimated. Additionally, the RMSE improves when increasing the Rician factor because of the greater dominance of the LoS component over multipath components.
We can see that our proposed approach outperforms the MUSIC algorithm in terms of RMSE. This is because unlike MUSIC that scans across predefined intervals to detect spectrum peaks, the CNN-based approach predicts sources' locations continuously and directly, enabling  finer resolution without relying on quantized grids. Additionally, by learning directly from data generated in multi-path environments, the proposed method is better equipped to handle challenging NLoS conditions. In contrast, MUSIC relies on the assumption that the received signal matches the array response structure, which typically holds only for the LoS component, leading to degraded localization accuracy in the presence of NLoS paths. More importantly, the computational complexity of the 3D MUSIC algorithm is prohibitively high, thus further hindering its application for real-time localization. 

\begin{figure*}[t]
	\centering
	\begin{subfigure}{0.325\textwidth}
		\centering
		\includegraphics[width=\linewidth]{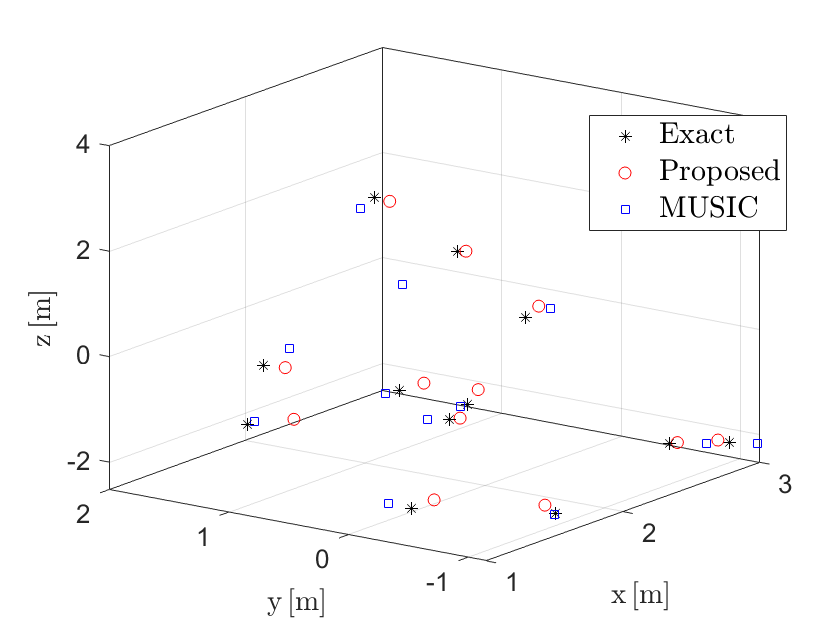}
		\caption{Source 1}
		\label{fig:source1}
	\end{subfigure}
	\hfill
	\begin{subfigure}{0.325\textwidth}
		\centering
		\includegraphics[width=\linewidth]{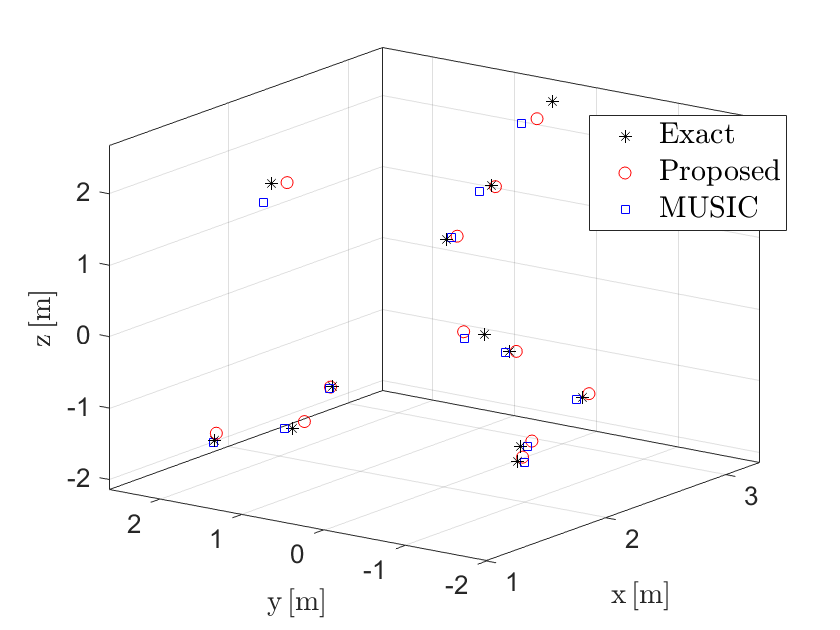}
		\caption{Source 2}
		\label{fig:source2}
	\end{subfigure}
	\hfill
	\begin{subfigure}{0.325\textwidth}
		\centering
		\includegraphics[width=\linewidth]{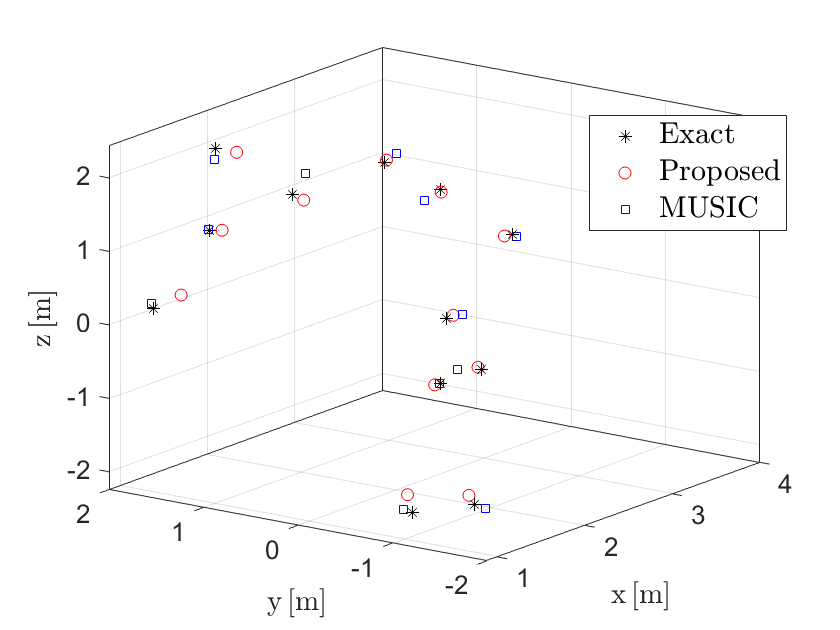}
		\caption{Source 3}
		\label{fig:source3}
	\end{subfigure}
	\caption{Exact and estimated locations for three sources over $12$ random realizations.}
	\label{fig:locations}

\end{figure*}

Fig.\,\ref{fig:locations} depicts the localization results for $12$ random realizations, where the black stars show the exact locations, red circles indicate the estimated locations via the proposed approach, and blue squares correspond to the locations obtained by MUSIC. In this simulation, the Rician factor is set as $\kappa = 4$, and $T = 25$ snapshots are used. We can see that our proposed approach achieves performance comparable to or better than MUSIC in most cases while requiring substantially less runtime once the network is trained, as discussed below. 

Table\,\ref{tab:RunTime} shows the runtime for the proposed approach and the MUSIC algorithm for different numbers of snapshots when $\kappa = 4$. We can see that there is a notable difference between the runtime of the proposed approach and the MUSIC algorithm. For example, for $T = 25$, our method achieves a runtime of just 0.0004 seconds, while it takes $99$ seconds for MUSIC to run. This represents an improvement of over $247000$ times faster execution.  Moreover, the runtime of the MUSIC algorithm increases proportionally with the number of snapshots, while our proposed approach maintains an almost constant execution time. The main reason for this observation is that MUSIC performs an exhaustive grid search and evaluates the spatial spectrum at every grid point, making it computationally intensive. In contrast, our trained model performs localization through simple matrix operations that have already been optimized during training, achieving significantly lower computational complexity.

\section{Conclusions}
This paper proposed a new  algorithm for the 3D localization of near-field sources with one anchor node using CNN. 
The CNN is trained using the eigenvectors of the covariance matrix to predict the 3D coordinates of near-field sources.
Simulation results show that the proposed approach outperforms the MUSIC algorithm while requiring less runtime, making it suitable for real-time localization applications.

\begin{table}[t]
\centering
\caption{3D localization runtime of the proposed algorithm and MUSIC algorithm for one test sample. The numbers represent the runtime in seconds.}
\begin{tabular}{|l||*{5}{c|}}\hline
\backslashbox{Method}{Snapshots}
&\makebox[2.8em]{$T = 25$}&\makebox[2.8em]{$T = 50$}&\makebox[2.8em]{$T = 75$}&\makebox[2.8em]{$T = 100$}
\\\specialrule{.25em}{.07em}{.075em}
Proposed algorithm & 0.0004  & 0.0004 &0.0004 &0.0005\\\hline
MUSIC algorithm &99 & 100& 104&127\\\hline
\end{tabular}\label{tab:RunTime}
\end{table}

\bibliographystyle{IEEEtran}
\bibliography{refs} 
\end{document}